 \documentclass[acmsmall]{acmart}

\def\BibTeX{{\rm B\kern-.05em{\sc i\kern-.025em b}\kern-.08emT\kern-.1667em\lower.7ex\hbox{E}\kern-.125emX}}

\copyrightyear{2019}
\acmYear{2019}
\setcopyright{acmlicensed}
\acmConference[CSCW '19]{CSCW '19: The 22nd ACM Conference on Computer-Supported Cooperative Work and Social Computing}{November 09--13, 2019}{Austin, Texas}
\acmBooktitle{Woodstock '19: The 22nd ACM Conference on Computer-Supported Cooperative Work and Social Computing, November 09--13, 2019, 2019, Austin, Texas}
\acmPrice{15.00}

\usepackage{graphicx}
\usepackage{subfig}
\usepackage{lipsum}
\usepackage{hyperref}
\usepackage{comment}
\usepackage{tabularx}
\usepackage{booktabs}
\usepackage{caption}
\usepackage{multicol}
\usepackage{comment}

 \usepackage{array}
 \usepackage{threeparttable}
 \usepackage{xcolor}
 \makeatletter
 \newcommand*{\@rowstyle}{}
\newcommand*{\rowstyle}[1]{
  \gdef\@rowstyle{#1}%
  \leavevmode\@rowstyle
  \ignorespaces
}
\newcolumntype{=}{
  >{\gdef\@rowstyle{}\ignorespaces}%
}
\newcolumntype{+}{
  >{\leavevmode\@rowstyle\ignorespaces}%
}

 \newcolumntype{C}[1]{>{\centering\arraybackslash}p{#1}}
 \makeatother

\begin{document}

%
\title{Your Stance is Exposed! Analysing Possible Factors for Stance Detection on Social Media}


%

\author{Abeer ALDayel}
\authornotemark[]
\email{a.aldayel@ed.ac.uk}
\affiliation{
  \institution{School of Informatics. The University of Edinburgh, Edinburgh, UK}
}
\author{Walid Magdy}
\authornotemark[]
\email{wmagdy@inf.ed.ac.uk}
\affiliation{
  \institution{School of Informatics. The University of Edinburgh, Edinburgh, UK}
}

%
\renewcommand{\shortauthors}{}

%
%
%

%

%

\begin{abstract}
To what extent user's stance towards a given topic could be inferred?  Most of the studies on stance detection have focused on analysing user's posts on a given topic to predict the stance. However, the stance in social media can be inferred from a mixture of signals that might reflect user's beliefs including posts and online interactions. This paper examines various online features of users to detect their stance towards different topics. We compare multiple set of features, including on-topic content, network interactions, user's preferences, and online network connections. Our objective is to understand the online signals that can reveal the users' stance. Experimentation is applied on tweets dataset from the SemEval stance detection task, which covers five topics. Results show that stance of a user can be detected with multiple signals of user's online activity, including their posts on the topic, the network they interact with or follow, the websites they visit, and the content they like. 
The performance of the stance modelling using different network features are comparable with the state-of-the-art reported model that used textual content only. In addition, combining network and content features leads to the highest reported performance to date on the SemEval dataset with F-measure of 72.49\%.

We further present an extensive analysis to show how these different set of features can reveal stance. 
Our findings have distinct privacy implications, where they highlight that stance is strongly embedded in user's online social network that, in principle, individuals can be profiled from their interactions and connections even when they do not post about the topic.
     
     
     
\end{abstract}

\maketitle
\textbf{This is a preprint of an article accepted for publication by CSCW 2019.}
\section{Introduction}
Stance towards a given topic is the position towards this topic either being in-favour or against it \cite{biber1988adverbial}.  Recently, large attention has been directed to automatic stance classification (detection) because of its wide range of applications, especially in the field of social media analysis. Earlier work focused on stance detection on argumentative debates in Online-forums \cite{somasundaran_recognizing_2009,murakami_support_2010,hasan2013extra}. With the wide spread of social media platforms, such as Twitter, which have become a common place for users to share their opinions towards various topics, research has been directed towards stance detection on these platforms. Detecting stance has widespread applications in social media analysis, opinion evolution, polarization detection, and rumours detection~\cite{garimella2017reducing,zubiaga_discourse-aware_2018}. Many studies used stance detection to analyze social media as main component of investigating the users aligns toward a given topic or
entity~\cite{mohammad_semeval-2016_2016,magdy__2016,benton2018using,lai2016friends,lai2018stance,AldayelMagdysocinfo}. 

Most research on stance detection modeled stance as a text classification task, where text of on-topic posts are used as the features \cite{mohammad_semeval-2016_2016,zarrella_mitre_2016,elfardy2016cu,siddiqua2018stance}. Some other work showed the effectiveness of using user's network as the features~\cite{darwish_improved_2017,magdy__2016,lai2016friends,lai2018stance}. However most of these studies were focused on one topic with no real examination to its generality on other topics or domains. 
Another limitation of the existing approaches for stance detection is the reliance on signals from active users only who frequently post on social media, where user's stance is modelled either by user's posts or interaction with other users (retweet in case of Twitter). There has been a growing interest on characterizing ``silent user'' in social media platforms ~\cite{bernstein2013quantifying,gong2015characterizing}. This group of users known as ''lurkers'' or ''invisible participants'' tends to contribute with a little or no content. Some users prefer to interact quietly on social media using other means of interactions instead of directly posting or sharing contents, such as following others and liking posts \cite{gong2015characterizing}. Most of stance detection studies used the network representation of the active users only and overlooked the silent users  ~\cite{darwish_improved_2017,magdy__2016,lai2018stance}.

\citeauthor{du2007stance}~\cite{du2007stance} argues that stance taking is a subjective and inter-subjective phenomenon in which stance-taking process is affected by personal opinion and non-personal factors such as cultural norms. Stance taking is a sophisticated process relates to different personal, cultural and social aspects. For instance, a political stance taking depends on experiential behavior as stated by
~\cite{mckendrick2014taking}. 
Thus users in social media might express their opinion directly by posting about the topic or their stance could be inferred indirectly through their interactions and preferences.
Our hypothesis is that user's embedded viewpoint in a post is related to the user's identity which could be better modeled by their interactions and connections in the social network. This idea is related to the concept of homophily in which users with same believes tend to have common interests and group together \cite{al2012homophily,garimella2018polarization,darwish_improved_2017}.

In this paper, we apply an extensive analysis to the possible online signals that can reveal the user's stance. To that end, we examine four groups of signals that might indicate the stance, namely: 1) on-topic posts by the user, which models users who explicitly express their stance on a topic; 2) user's interactions on social media with other users or websites, which models users interactions online regardless having them expressing their stance or not (IN); 3) user's preferences the posts they like, which enable modeling silent users who do not post or share content only (PN); and finally 4) the network of users they are connected, which enable modeling passive users who might have no content or interaction on social media, but just follow other accounts online (CN). We compare the effectiveness of each of these groups of features on detecting stance individually and when combined. Our main research question is to understand ``What are the factors that can reveal the stance of user online towards a given topic''. We further analyse ``how'' and ``why'' these factors might be effective  for detecting stance. Our list of research questions in this paper are:
\begin{itemize}
    \item What are the different signals in user's online activity that can reveal their stance, including textual content, networks of interaction (IN), preference (PN), and connection (CN)?
    \item Does the performance of detection differ by different types of topics?
    \item What makes any of these signals effective (or ineffective) for detecting stance?
\end{itemize}

Our experiments are applied on the SemEval stance detection benchmark dataset~\cite{mohammad_semeval-2016_2016}, which contains a set of over 4,000 tweets labeled by stance towards five different topics. The five topics covers multiple domains not just politics, which makes the dataset ideal to examine the generalisability of the stance detection models, unlike most of work in literature that typically focus on studying one political topic at a time~\cite{lai2016friends,lai2018stance,magdy__2016,darwish_improved_2017}. Our results show that training a classification model on pure user network features outperforms the state-of-the-art baseline system~\cite{mohammad_semeval-2016_2016} which is trained on multiple features extracted from the tweets text content. This includes when using the preference network features from only the tweets the user likes and also the connection network of the accounts the user follow, where both can model silent users.
When different groups of features are combined, including content and network, a significant improvement is observed. Our findings suggest that for the task of stance detection, even when applied on the level of tweet, user's network information are more effective features than the content of the tweet itself. This aligns to the sociolinguistic theory in \cite{bassiouney2015stance}, where it defines stance as the link between linguistic forms and social identities which has the capability to establish the alignment between stance-takers. 

We further applied an extensive analysis to the most influential features for each group of network signals to understand how they outperform textual text. It was interesting to find that the overlap between IN, PN, and CN was not large, where the common nodes among them are around 10\% only, however, each of those networks still can model user's stance towards a given topic. Our analysis to the most influential features from each network on each of the five topics shows that there is usually some common signals in user online activity that can reveal their stance towards a given topic regardless of the type of the topic. We believe that our findings in this study raises a large concern about protecting the privacy of social media users, where their beliefs and leanings could be easily predicted using any of the footprint signals they leave online. This should motivate social media networks owners and designers to develop methods for protecting the privacy of their users~\cite{waniek2018hiding}.

The collected network information for the SemEval dataset would be made publicly available to allow replication to our experimentation\footnote{\url{https://github.com/AbeerAldayel/Stance_detection}}.

\section{Related work}

There is a considerable amount of work on viewpoint or stance detection; yet, less work compared the role of content and social actor interactions in stance detection \cite{himelboim2013birds,darwish_improved_2017}. Studying stance needs to cover the intersection dimensions of stance taking process, which are mainly influenced by linguistic forms and social interactions frames~\cite{mckendrick2014taking}. Most of the previous studies define stance as a textual entailment task where the main processing depends on the raw text only \cite{dey2018topical,mohtarami2018automatic,augenstein2016usfd,mohammad_semeval-2016_2016,rui_weakly-guided_2017}. In this form of stance detection, a given text entails a stance towards a premise (target). 

It has been shown that constructing a knowledge based dataset about the topic is beneficial in stance detection task \cite{mohammad_semeval-2016_2016}. This constitutes a visible hurdle which limits the stance detection task to set of predefined topics. Furthermore, many times the topic is not mentioned in the tweet. 
One way that was suggested to handle the unmentioned target entity in text is to analyze the opinion to the opponent of the entity or supporter of the entity.
For example, \cite{dias_inf-ufrgs-opinion-mining_2016} constructed a list of keywords that identifies Trump using a dataset labeled with stances toward Hillary. Using this list of keywords help in detecting the unexpressed stand towards Trump. Another study \cite{rui_weakly-guided_2017} follows the same line by constructing corpus that contains words that are \textit{against} and \textit{in-favor} each target to enrich the models. Similarly, 
\cite{wei2016pkudblab} used a domain corpus related to Trump along with lexicon to construct a labeled dataset to detect stance towards Trump. Furthermore, \cite{benton2018using} used context of the users tweets to construct author embedding and predict the stance. 

There has been some work on studying the integration of network and content with a limited focus on the ideological political views  \cite{darwish2019unsupervised,himelboim2013birds,lai2016friends,magdy__2016}. For instance the study of \cite{himelboim2013birds} focused on the liberal and conservative on twitter. 
Unlike previous work, rather than studying the stance on single topic and using a domain specific data, we study the stance in various domains. This study explores the stance modeling in the social media to know to what extent do network interactions and content interactions reveal an individual's viewpoint. Examining the implications of those interactions in detecting users' stances provides a better understanding of stance modeling on social media.  
In the following we summarise the literature work on stance detection.


\subsection{Stance Detection on Twitter}

The task of detecting stances takes a way back, focusing on online debates in online forums
\cite{lin2006side,murakami_support_2010,walker_corpus_2012}. With the widespread of social media, they soon become a rich source of argumentative data, which has attracted many researchers to study stance detection on these platforms. As these platforms foster the real-time engagements with the new events, many studies used data collected from social media to predict people stances towards different topics \cite{gu2017ideology,zeschcomparing,magdy__2016}. For instance, the study done by \cite{zeschcomparing} designed a stance detection model using YouTube's comments data. 

Over the last decade, Twitter has become the most commonly used  platform to study the expressed stance towards various events/topic \cite{djemili2014does,magdy__2016,lai2016friends,fraisier2018stance}. This platform featured to be open and capable to reach a significant proportion of audience. As a social media platform, the network structure in Twitter has a profound existence through various features provided within this platform. These features have made Twittersphere an attractive source of data to study stance and detect opinions toward a broad range of topics in real-time. 
In this platform users can connect and interact with each other directly through reply, retweet or mention. The retweet interactions considered a asymmetric interaction. In this kind of communication the user can retweet a tweet without the author acknowledgment.  In contrast, the reply takes a form of symmetric communication where both users are involved in the process of interaction \cite{mejova2015twitter}. Basically, each user has a home timeline shows a stream of Tweets from accounts the user have followed on Twitter. Within this home timeline, user can reply, retweet, or like a Tweet from within the timeline. The collection of liked tweets for each user is shown in ``Likes'' (sometimes referred to ``Favourite'') timeline. The Likes timeline include only public liked tweets. If a user liked a ``private'' tweet of protected account they follow, it will not show up in their Likes timeline. Beside the endorsement and interactions, Twitter users can have a set of ``Followers'' and ``Friends''. The ``Friends'' collection contains accounts that a specific Twitter user follows.
The Friends and Followers networks have been used effectively in previous studies to capture the social ties ~\cite{butts2009revisiting,xie2012friend}.

\subsection{SemEval Stance Detection Task}
One of the well known stance dataset derived from Twitter is the SemEval stance dataset. This dataset is designed for supervised stance detection (task A) \cite{mohammad_semeval-2016_2016}. The dataset contains a (topic,tweet) pair for five topics covering political, social, and religious domains. Over 4000 tweets are released in this dataset, each labeled with stance as {favor, against, or none} to one of the five topics.

19 teams have participated in the task and submitted different models for stance classification.
Most of the participants developed models that learn linguistic cues from the given tweet text to identify the stance for the target~\cite{mohammad_semeval-2016_2016,elfardy2016cu}. Others used text representation methodologies such as LSTM conditional encoding to represent tweet-target pairs~\cite{augenstein_stance_2016}. One team, MITRE \cite{zarrella_mitre_2016}, obtained a result with overall F-score of 67.82\% by using two recurrent neural networks (RNN) classifiers. Another team, Pkudblab \cite{wei_pkudblab_2016}, achieved 67.33\% F-score by utilizing convolutional neural network (CNN).
While most of these approaches used various methods for creating an effective stance classifier, at the time of the competition, the best reported system used a simple character and word n-grams representation for the tweet text to train a linear SVM model, which achieved an average F-score of 68.98\% \cite{mohammad_semeval-2016_2016}. 

Several studies have been published later on the same SemEval dataset reporting similar or marginal improvements to the performance while continuing to use various representations of the text to train different machine learning models. \cite{du2017stance} proposed attention-based neural network by using a target-specific information which produced an overall F-score of 68.79\%. 
Another marginal improvement introduced by \cite{dey2018topical} by using attention based LSTM model, which achieved 68.84\% F-score. Whereas in \cite{zhou_connecting_2017} the usage of bi-directional GRU-CNN yielded an F-score of 69.42\%.

To the best of our knowledge, the current highest reported performance on this dataset is by ~\cite{siddiqua2018stance} who trained multiple SVM models on only two classes (against and favor), while neglecting the ``none'' class. Forcing the classifier to predict a polarised stance led to the highest reported result on this benchmark dataset with F-score  of 70.03.

\subsection{Network Features to Detect Unexpressed View}

Another stream of research on social media analysis has shown the large split in the networks of online social media~\cite{garimella2017reducing,garimella2018quantifying}. This has been analyzed as a reason to the social phenomenon of homophily~\cite{al2012homophily,bessi2016homophily} that states the fact of users with similar believes tend to interact with each other, which creates what is so-called echo-chambers in online social networks~\cite{colleoni2014echo,guo2018responsible}. Few studies have utilized this phenomenon as a feature to predict unseen views of social media users on topics that they never discussed~\cite{magdy__2016, darwish_improved_2017,lai2018stance}.  This assumption is powered by~\cite{jaffe2009stance} in which she describes stance as non-transparent act in the text and must be inferred from the empirical study of interactions. 
Different sets of user features have been introduced in the previous studies with the focus of defining similar users for specific events.
For instance \cite{magdy__2016} utilized user-interaction to predict users who would share hate-speech against Muslims after Paris attacks in 2015 using other accounts that a user mention, retweet, and reply to. They reported that using user's network interactions can predict the user's future attitude towards Muslims with an accuracy of 88\% even when the user never discussed any topic related to Muslims before.
Another work by \cite{darwish_improved_2017}, proposed a user similarity measure that is based on the distance between users in a social network to predict the users' stance towards two different topics. Similarly \citeauthor{lai2016friends} integrated network features to study users political leaning in the US elections 2016 and the Italian political debates~\cite{lai2018stance}.

Further work studied the users interactions as a factor to model the stance in the social media. The work of \cite{trabelsi2018unsupervised} used  graph partitioning method with social interactions to cluster the users based on their viewpoint. Furthermore the study of \cite{thonet2017users} used the interactions between users with focus on the retweet and reply network as way to cluster the users with the same views. Similarly, \cite{fraisier2018stance} used the users interactions and the textual features to model the users stance by proximity graphs.

These studies highlights the importance of social network interaction of users to detect their position towards specific events or entities. Nevertheless, they are limited to focusing on one specific topic from the political domain, which lacks examining the generalisability of these approaches on multiple topics from different domains. In addition, it focuses on network interactions that can only model active users who retweet, mention, and reply other accounts.

In this paper, we utilise the SemEval benchmark dataset to apply an extensive comparison on stance detection using multiple sets of features and compare it to the state-of-the-art.  
In addition, we introduce the use of the preference network as a new way to model the stance and examine the possibility of detecting the stance of the silent users. We compare the performance of this new set of feature with content-based and other networks based features for stance detection.

\section{Stance Detection Methodology}
In the following, we discuss our proposed methodology including the set of features used and the machine learning method applied. But initially, we discuss the implications of our approach from the conceptual point of view.

\subsection{User vs Tweet Level Stance Detection}
The SemEval dataset is labeled for stance on the tweet level, while we are examining user-level features. To enable comparison to state-of-the-art methods on the same dataset, we apply our detection on the tweet level. This would not be an issue if each tweet in the dataset is coming from a different users. However, we noticed that 167 users (out of 3,528) in the dataset produced multiple tweets. 
This means that our classifiers trained on network features would always give the same classification to any tweets posted by the same user. We argue that this is acceptable based on the assumption that user stance for a given topic is not expected to change within a short period of time \cite{borge2015content}. Moreover, we hypothesize that even if the user stance gets changed, it would be accompanied by a change in the network interactions of the user \cite{waniek2018hiding}. 

To further validate our assumption, We examined the 167 users who produced multiple tweets on the same topic. Out of those, 104 users have fixed stance in their multiple tweets and 42 have fixed polarized stance with some tweets with no stance (labeled \textit{none}). Only 19 users have a mix of \textit{favour} and \textit{against} stance in the same topic, but with clear dominance for one of them (e.g. 16 vs 1 tweets). This quick analysis, shows that the majority of tweets from the same user are expected to have a fixed stance on a single topic. Thus, we believe that having a fixed set of features, based on user's network, for all tweets of the same user can be seen as an acceptable approach for stance detection on the tweet level.

\subsection{Feature Extraction}

We define four features sets to model the stance in social media. These sets are:  on-topic content, user's network interactions, preferences and connections. Those are defined as follow:
\begin{itemize}
    \item \textbf{On-Topic Content (TXT)}, models the text of the tweet, including features combining both word and character n-grams as presented in the best performing system in SemEVal 2016~\cite{mohammad_semeval-2016_2016}. This set of features models stance of users who explicitly express it in text. 
    \item \textbf{Interaction Network (IN)}, models the network the user interacts with in their posts. It includes the mentioned accounts ($IN_{@}$) and website domains ($IN_{DM}$) the user interacts with directly either by retweeting, replying, mentioning, or linking. 
    \item \textbf{Preference Network (PN)}, models the network the user prefers from the tweets they like. It includes the mentioned accounts ($PN_{@}$) and linked website domains ($PN_{DM}$) in the tweets the user likes. 
    \item \textbf{Connection Network (CN)},  models the online social ties between the users, which includes the accounts who follow the users (followers $CN_{FL}$), and those the user follows (friends $CN_{FR}$). 
\end{itemize}



Table \ref{tab:Features} shows a detailed explanation of the feature sets. It's worth noting that $IN$ features are independent of having users expressing their stance towards the target topic, since it depends on the social and web networks the user interact directly with regardless to the content in tweets. Both $PN$ and $CN$ features enable modeling silent or passive users who do not post or share content rather than just following or liking tweets from others.
Our objective is to understand how each of these feature sets would compare to each other and to the textual features which have been studied heavily in literature.





\begin{table}
\centering
\small
\begin{tabular}{ll} 
  \hline
  \textbf{Feature Set}  & \textbf{Description} \\
   \hline
   
  TXT & word and character n-grams of the tweet text. \\ 
  IN: & user's interaction network. Extracted from user's \textit{Home} timeline.\\
  - IN$_{@}$ & the list of accounts the user retweet for, reply to, or mention in their timeline. \\
  - IN$_{DM}$ & the list of web domains the user link in their tweets.\\
  PN: & user's preference network. Extracted from user's \textit{Likes} timeline. \\
  - PN$_{@}$ & the list of accounts mentioned in the tweets the user likes.\\
  - PN$_{DM}$ & the list of web domains in the tweets the user likes. \\
  CN: & user's connection network. Accounts user connected to.\\
  - CN$_{FL}$ & the list of followers of the user, i.e. accounts that follow the user. \\
  - CN$_{FR}$ &  the list of followees/friends, i.e. accounts that the user follows. \\
 
  \hline
\end{tabular}
\caption{List of feature sets examined in our experiments with their description.}
\label{tab:Features}
\end{table}

\subsection{Stance Detection Model}
Since our main contribution is on stance representation to analyse the effectiveness of different social signals in detecting stance, we used our proposed set of features to train an SVM model with linear kernel for two main reasons: 1) It achieved the best performing model over 19 participating groups at SemEval 2016 ~\cite{mohammad_semeval-2016_2016} while outperforming more sophisticated model that used deep learning~\cite{augenstein_stance_2016,zarrella_mitre_2016,wei_pkudblab_2016}. 2) SVM models built with linear kernel are easily to interpret, which would enable us to apply feature analysis for a better understanding to the influential features and their role in stance detection.
We used Scikit-learn\footnote{Scikit-learn\url{http://sciki-learn.org/}} implementation of SVM, which use cross-validation with k=5. 
For all the features, we use vector representation with Boolean value to indicate the presence or absence of the feature's values. We have examined other feature values, such as frequency and tf-idf, but Boolean values showed the best performance.

\section{Experimental setup}

\begin{table*}[t]
\centering
\begin{tabular}{lllll}
  \hline
  &  \multicolumn{2}{c}{\textbf{Full dataset}} &  \multicolumn{2}{c}{\textbf{Existing Users}}\\
  \textbf{Topic} & Train & Test & Train & Test \\
  \hline 
     Atheism (A) &  513 (434) & 220 (196)& 380 (302) & 170 (146) \\
     Climate change is a real concern (CC) &  395 (347) & 169 (145) & 317 (269)& 144 (120)\\ 
     Hillary Clinton (HC) &  639 (556) & 295 (250) & 447 (364) & 223 (178) \\ 
     Feminist movement (FM) &  664 (620) & 285 (256)& 354 (312) & 170 (141) \\ 
     Legalization of abortion (LA) & 603 (496) & 280 (228) & 471 (365)& 199 (147)\\
     \hline
     Total & 2814 (2453) & 1249 (1075) & 1969 (1612) & 906 (732) \\
  \hline
\end{tabular}
\caption{Number of tweets used for training and testing with respect to Semeval 2016 topic. The number of unique users authored the tweets are shown in brackets.}
\label{table:data_users}
\end{table*}

\subsection{Data Collection}

Our experimentation has been applied to the benchmark dataset of the SemEval 2016 stance detection task \cite{mohammad_semeval-2016_2016}. The dataset contains a set of 2814 and 1249 tweets for train and test respectively covering five topics. These topics are: Atheism (A), Climate Change (CC), Feminist Movement (FM), Hillary Clinton (HC), and Legalisation of Abortion (LA). As could be noticed, these topics are not just political, but actually covers topics of social (e.g. `FM', `LA') and religious (e,.g. `A') natures.

We further used the Twitter REST API to collect the network information of the users in SemEval stance dataset. Basically, we collected two timelines for each of the users posted the tweets in our dataset, namely \textit{Home} timeline\footnote{\url{https://developer.twitter.com/en/docs/tweets/timelines/api-reference/get-statuses-home_timeline.html}}, which we use to contruct the user's IN; and the \textit{Likes} timeline\footnote{\url{https://developer.twitter.com/en/docs/tweets/post-and-engage/api-reference/get-favorites-list.html}}, which we use to construct the user's PN. In addition, we collected the user's list of followers and friends to construct the user's CN\footnote{\url{https://developer.twitter.com/en/docs/accounts-and-users/follow-search-get-users/overview}}.
Unfortunately, we found that around 25\% of these users have been deleted or suspended. Therefore, we end up with smaller number of tweets in the collection that we can apply our approach to them, exactly 1969 training and 906 testing data\footnote{List of ids of tweets and users network information would be made available}. 
Table \ref{table:data_users} shows the distribution of tweets (and users authored them) that we could retrieve in our dataset compared to the original SemEval dataset.


For each of the users in our collection, we managed to collect an average of 2,552 and 1,801 tweets from the \textit{Home} and \textit{Likes} timelines respectively. For each user, the set of mentions in those timelines were extracted and saved separately. In addition, we collected the set of friends and followers of each of the users in our collection. $IN_{@}$, $PN_{@}$, $CN_{FR}$, $CN_{FL}$ represent the set of unique accounts appeared in the user's tweets, likes timelines, list of friends, list of followers of the user respectively.
In addition, all the links appeared in the timelines were extracted and expanded (in case they were shortened). The domain of each link was then extracted and saved. $IN_{DM}$ and $PN_{DM}$ represent the set of unique web domains appeared in the user's tweets and likes timelines respectively.
\subsection{Baselines and Evaluation}

We created two baseline systems that achieve the highest reported performance on the SemEval dataset based on the best performing participating system in the SemEval task~\cite{mohammad_semeval-2016_2016} that is trained on the \textit{TXT}\space  features. 
For the first baseline, an SVM with linear kernel trained on the three stance classes using a combination of both word and character n-grams was used to represent the textual content of the tweet to be classified. Word n-grams was used with $n$ = \{1, 2, 3\}, and character n-grams was used with $n$ = \{2, 3, 4, 5\}. These features were used to train the SVM classifier with linear kernel. We only used the subset of training data that we managed to retrieve its users network information to allow a direct comparison to our models. The outcome of this model achieved an average F-score of 68.48 on our subset of the test data, which should be comparable to the reported best model in~\cite{mohammad_semeval-2016_2016} that achieved an average 68.98 F-score but on the whole dataset.

From a sociolinguistic perspective, it has been argued that there is no complete neutral stance as people use to position themselves with favor or against the object of evaluation \cite{jaffe2009stance}. To comply with this argument, we created our second baseline by retraining the same SVM classifier with the same set of features, but with considering only the two polarised classes \{favor, against\} and neglecting the `none' class. In this way, we force classifier to have a decision on the polarised stance of the user. While this approach will misclassify the samples in the test set with ground-truth `none' stance, it was shown in the current state-of-the-art system~\cite{siddiqua2018stance} that this approach actually outperform the three-class classifier, where they achieved F-Score of 70\% when trained a binary SVM classifier with tree kernel after neglecting the `none' class. When we applied this approach, the overall F-score of the system got an actual improvement to reach 69.8\%, which is comparable to~\cite{siddiqua2018stance}.



After building the linear SVM baselines (both with the three and binary classes models), we trained the same models with the different set of suggested network features. We test each feature set separately and compare their performance to the models that depends on the tweet textual content; then we apply different combination of the features to observe any potential improvement in the performance.

To evaluate the performance of our method, we used the official SemEval-2016 macro-average of the F1 score for the `Against' and `Favour', where the F-score on the `None' class is discarded from calculating the average \cite{mohammad_semeval-2016_2016}. The same evaluation script provided by SemEval stance detection task was used to report the results.
In addition, we show the performance over each of the five topics separately for a deeper analysis of the performance. 

\section{Results}

\begin{table}
\centering
\small
\begin{tabular}{l|rrrrr|rrr} 
  \hline 
  & &  & \textbf{Topic} & & &  & \textbf{Overall} &  \\
  \textbf{Model}  & A &CC &HC &FM &LA  & F$_{favour}$ & F$_{against}$ & F$_{avg} $   \\
  \hline
  TXT (Baseline)  &61.38 & 42.86 & 58.91 & 52.01 & 60.96  & 63.09 & 73.87 & 68.48 \\
  
\hline
  IN$_{@}$  & 68.94 & 40.09 & 62.15 & 54.80 & 56.25 & 60.77& 75.57& 68.17 \\
  IN$_{DM}$  & 56.86 & 38.46 & 34.20 & 38.67 & 53.31   & 49.19& 61.76 & 55.47\\
  IN$_{@}$+IN$_{DM}$ & 70.16 & 39.81 & 61.59 & 57.63 & 64.16  & 64.04 & 76.18 & \textbf{70.11}  \\
  \hline
  PN$_{@}$  & 73.30 & 36.36 & 56.82 & 48.43 & 56.41  & 55.81 & 73.39 & \textbf{64.60} \\
  PN$_{DM}$  & 62.99 & 35.18 & 58.01 & 46.71 & 48.49  & 50.85 & 70.26 & 60.56 \\
  PN$_{@}$+PN$_{DM}$  & 64.55 & 37.13 & 54.27 & 49.00 & 56.44  & 55.73 & 70.14& 62.94\\
  
  \hline
 
CN$_{FR}$ & 66.71 & 30.11 & 63.87 & 51.51 & 53.10 &  51.15&72.76 & \textbf{61.96}\\
CN$_{FL}$ & 40.78 & 20.29 & 54.11 & 46.80 & 56.38 & 39.55 & 65.82& 52.68\\
CN$_{FR}$+CN$_{FL}$ &  49.66 &28.14&  66.95& 48.76 & 49.72 & 44.85 &  67.98&  56.42\\
  \hline
  
\end{tabular}
\caption{Stance detection performance using different set of features using SVM classifier trained on three classes. F-Score (\%) is reported on the SemEval stance detection task for each topic and overall.}
\label{tab:results}

\begin{tabular}{l|rrrrr|rrr} 
  \hline 
  & &  & \textbf{Topic} & & &  & \textbf{Overall} &  \\
  \textbf{Model}  & A &CC &HC &FM &LA & F$_{favour}$ & F$_{against}$ & F$_{avg} $   \\
  \hline
  TXT (Baseline)  &61.91& 42.86& 59.53&	52.21&	62.40 & 63.53& 76.07 & 69.80 \\
  \hline
  IN$_{@}$  &68.30 	& 54.14	& 59.05	& 50.40	& 60.82 & 61.89 & 77.90 & 69.89 \\
  
  IN$_{DM}$   &63.24& 	42.86&	53.91&	61.24 & 60.17 & 61.51 & 76.82 & 69.17 \\
  
  IN$_{@}$+IN$_{DM}$ & 67.65&	 42.86 & 62.64 & 55.87 &	63.93 & 64.04 & 79.07 &\textbf{71.56} \\
  \hline
  PN$_{@}$  & 73.49 & 42.86 & 59.26 & 49.63 & 63.87 & 63.70 & 77.70 & 70.7 \\
  PN$_{DM}$ &67.14	&42.17	&58.33	& 51.62 & 61.79& 60.18& 77.28 & 68.73  \\
  PN$_{@}$+PN$_{DM}$  &68.03&	42.86&	59.00 &	52.57&	65.50&  63.91& 78.60 &\textbf{71.25}\\
  
  \hline
 CN$_{FR}$ & 63.83	& 42.86	& 64.01	& 60.93	& 59.58& 64.53 & 78.25& \textbf{71.39}  \\
  
  CN$_{FL}$ & 35.97	& 	42.86& 58.51	& 52.70	&62.68 & 56.08 &69.73 & 62.91 \\
  
 CN$_{FR}$+CN$_{FL}$& 50.00	& 42.86	& 68.21	& 57.38	& 54.13&  58.07& 73.41& 65.74 \\
  \hline
  
\end{tabular}
\caption{Stance detection performance using different set of features using \textit{binary} SVM classifier. F-Score (\%) is reported on the SemEval stance detection task for each topic and overall.}
\label{tab:results_two_Classes}
\end{table}



\subsection{Stance Detection Results}
Tables \ref{tab:results} and \ref{tab:results_two_Classes} report the performance of the three-class classifier and binary classifier for stance detection, respectively. 
The general observation from the tables is that the binary classifier outperforms the classifier that is trained on three classes. While the binary classifier misclassifies tweets with no stance, it is more effective in detecting the polarised stance. This initial observation shows that forcing automatic classifiers to decide on a given stance might be a more effective approach than allowing them to have the `none' option about stance, which makes it more confusing following \citeauthor{jaffe2009stance}'s argument that there is no complete neutral stance~\cite{jaffe2009stance}. We analyse this further in the following subsection.


The second observation, for the binary classifier (Table~\ref{tab:results_two_Classes}), is that all the three set of network features - that are totally independent of the tweets contents - have better overall performance than the state of the art systems that depend on tweets textual content.
In fact, the activity network ($IN$) and the preference network ($PN$) features that combine the accounts and domains features achieve better results than the baseline on all the five topics. This confirms the consistent performance of network features over text on topics of different domains.
In the connection network ($CN$), the friends network ($CN_{FR}$, the accounts the user follows) outperformed the baseline, while the follower network ($CN_{FL}$) achieved the lowest average F-score among all classifiers, even when combined with the friends network. This is potentially because of the sparsity of this network, where finding common followers among different users is less likely compared to finding common accounts they might follow, where it is expected to have people of similar stance following common accounts as a part of the homophily phenomena in social media~\cite{al2012homophily,garimella2018polarization}.

While user's interaction network showed the best overall performance among all feature sets, Table \ref{tab:results_two_Classes}, it was interesting to see preference network outperformed all models in two of the five topics when using the binary classifier. These results support the hypothesis about stance detection, which is the online social network activity of a user posting a tweet contains enough signals to detect the stance of tweet regardless of its content. Furthermore, we show that the preference network of user's likes on Twitter still can achieve decent detection of stance, which enables detecting stance for silent users.

\begin{table}
\centering
\begin{tabular}{lrrr} 
  \hline
  \textbf{Model}  & F$_{favour}$ & F$_{against}$ & F$_{avg} $  \\
  \hline
  (A) TXT+IN$_{@}$+IN$_{DM}$   & 67.21 & 76.49 & 71.85 \\
  (B) TXT+IN$_{@}$+IN$_{DM}$ & 66.67 & 78.31 & \textbf{72.49} \\
  \hline
\end{tabular}
\caption{The result of baseline linear SVM model when combining both text and network features. Model (A) and (B) shows the result when trained on three and two classes, respectively.}
\label{tab:resultsall}
\end{table}

We further tested combining the best performing network features from the two networks (IN$_{@}$+IN$_{DM}$) with  (TXT) to see if this can further improve the performance. Table~\ref{tab:resultsall} shows the best achieved average F-score when we combined the network with content features, where the best performance achieved when we combined the interaction network with text for both the three-class and the binary classifiers\footnote{we also tested other combinations of feature sets, but TXT+IN$_{@}$+IN$_{DM}$ achieved the highest results}. This result was found to be statistically significantly better than the state-of-the-art baseline model using two-tailed t-test with $p-value$ $<$ 0.05 (we also tested significance using Mann-Whitney U test \cite{mcknight2010mann}, but it did not show significance).



\subsection{ Performance Discussion}

\begin{figure}
  \subfloat[Three Classes IN$_{@}$ + IN$_{DM}$]{
	\begin{minipage}[c][1\width]{
	   0.3\textwidth}
	   \centering
	   \includegraphics[width=1\textwidth]{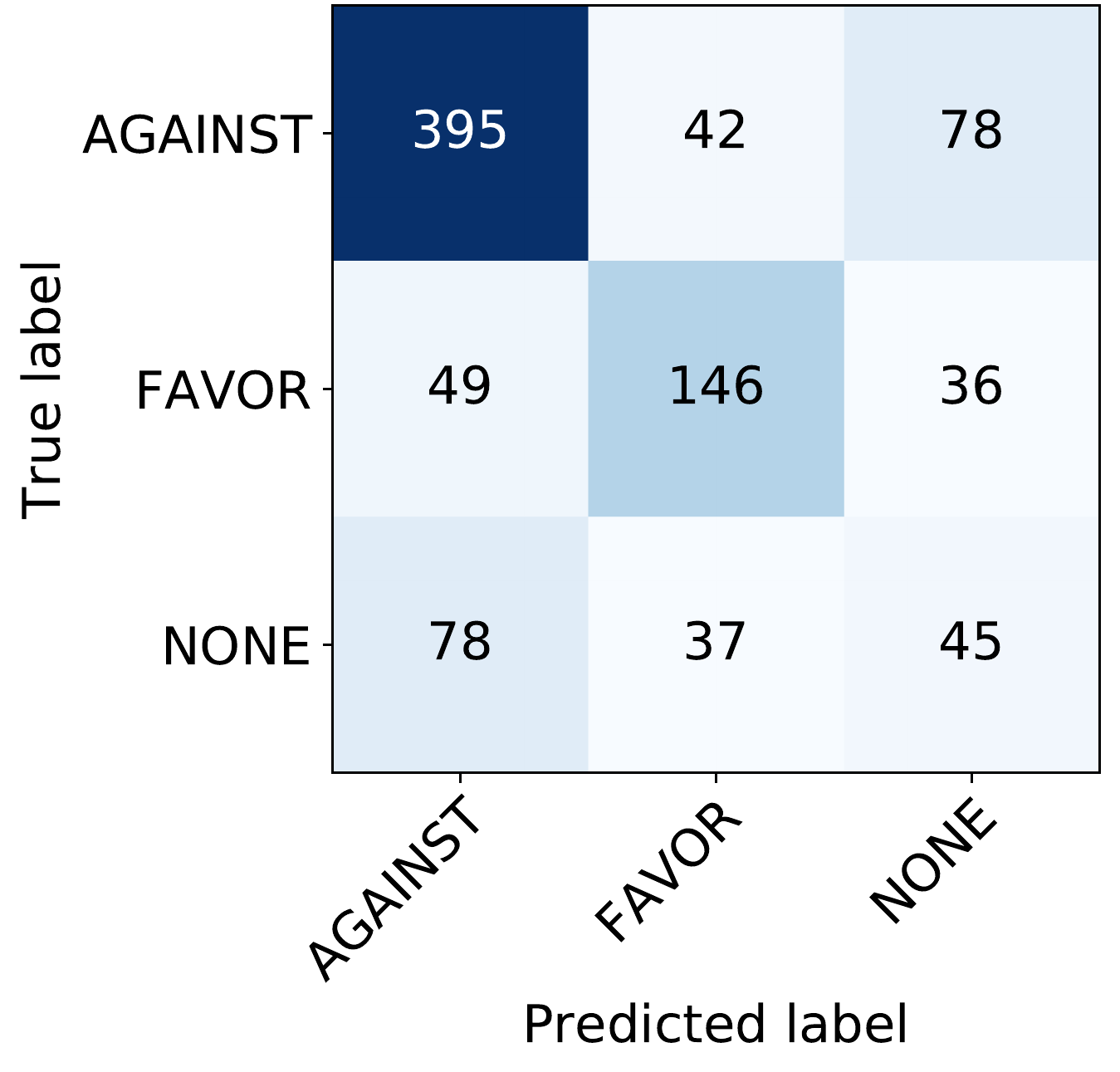}
	\end{minipage}}
	\hspace{0.5cm}
  \subfloat[Two classes IN$_{@}$ + IN$_{DM}$]{
	\begin{minipage}[c][1\width]{
	   0.3\textwidth}
	   \centering
	   \includegraphics[width=1\textwidth]{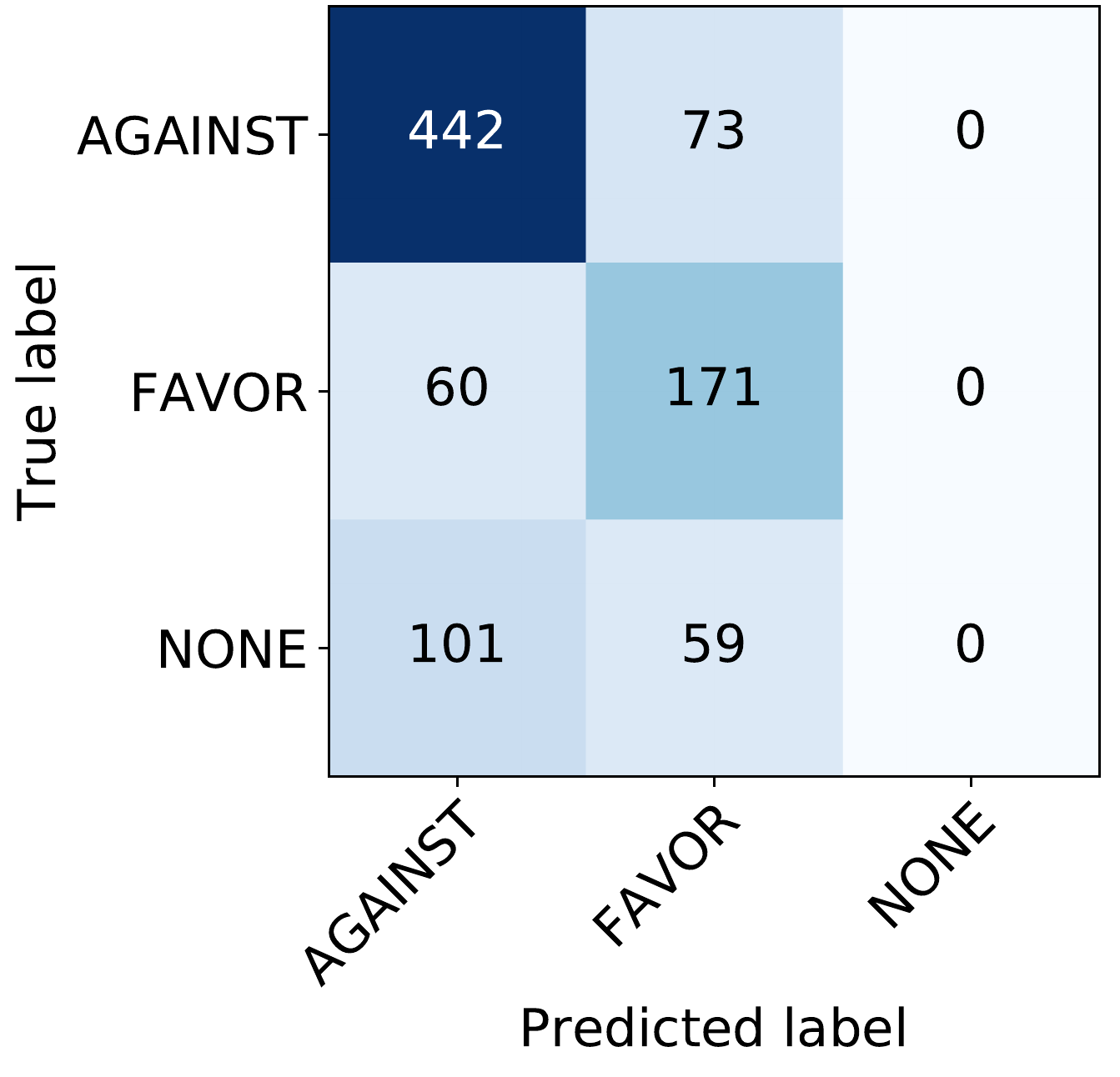}
	\end{minipage}}
\caption{Confusion matrices for the best three vs two classes prediction models. }
\label{fig:conf_Best_models_new}
\end{figure}

As shown earlier, forcing the stance model to predict in-favor and against stances and ignore the `none' stance consistently leads to better performance using all feature sets. This is an interesting result, since a binary classifier will always misclassify the `none' class leading to larger number of false positives to the other two main polarised classes, which should reduce the performance. To better understand this, we plot the confusion matrices for the best performing model for both three/two class classifiers in Figure \ref{fig:conf_Best_models_new}. As it is shown, the binary classifier led to larger number of false positives for both the polarised classes; however at the same time, it led to larger number of true positives for both classes. This led to improvement in recall with some reduction in precision, with an overall improvement in the average F-score. 

Another observation from Tables \ref{tab:results} and \ref{tab:results_two_Classes}, is the low performance of classifying stance on the climate change (CC) topic, where it has the lowest F-score among all topics. We conducted a further analysis and we noticed a large difference in the class distribution between the `in-favor' and `against' classes, where 176 samples in the training set are labeled as `in-favor', while only 8 samples are labeled as `against'. This led the classification models to predict the majority class in most of the cases, which led to random-like performance for this topic.



Our obtained results for stance classification are the highest to be reported to date on the SemEval dataset, which confirms the large impact of utilising user's network activity as features in boosting the performance of stance detection, especially when combined with textual features. Our results highlight that user's stance towards given topics could be inferred with various types of features from their activity online. In the following section, we apply an extensive analysis to these features to understand its role and influence in revealing the user's stance.

\section{Feature Analysis}
In this section, we analyse each of the network features that showed to be effective in detecting stance. We apply our analysis to the binary classifier, which achieved the highest results. Our analysis includes studying differences between our three networks, analysing most influential features per network and per topic, and giving examples of how these features might be effective.

\subsection{Similarity between Networks}

From the results obtained in Table~\ref{tab:results_two_Classes}, it is noticed that the scores achieved by the three groups of networks (IN), (PN) and (CN) are relatively similar. The average F-scores obtained by (IN$_{@}$ + IN$_{DM}$), (PN$_{@}$ + PN$_{DM}$) and (CN$_{FR}$) are around 71\%  and their results were found to be statistically indistinguishable from each other using both t-test and Mann-Whitney U test. This motivates to further examine the overlap among these networks, since it is highly possible that users interact with and like content of the same set of accounts they follow.
Hence, we measure the overlap between the features of (IN), (PN) and (CN) to gauge the similarity among them.

\begin{figure*}
    \centering
    \setkeys{Gin}{width=0.40\textwidth}
\subfloat[ IN$_{@}$-PN$_{@}$-CN$_{FR}$.
          \label{fig:subfig-a}]{\includegraphics{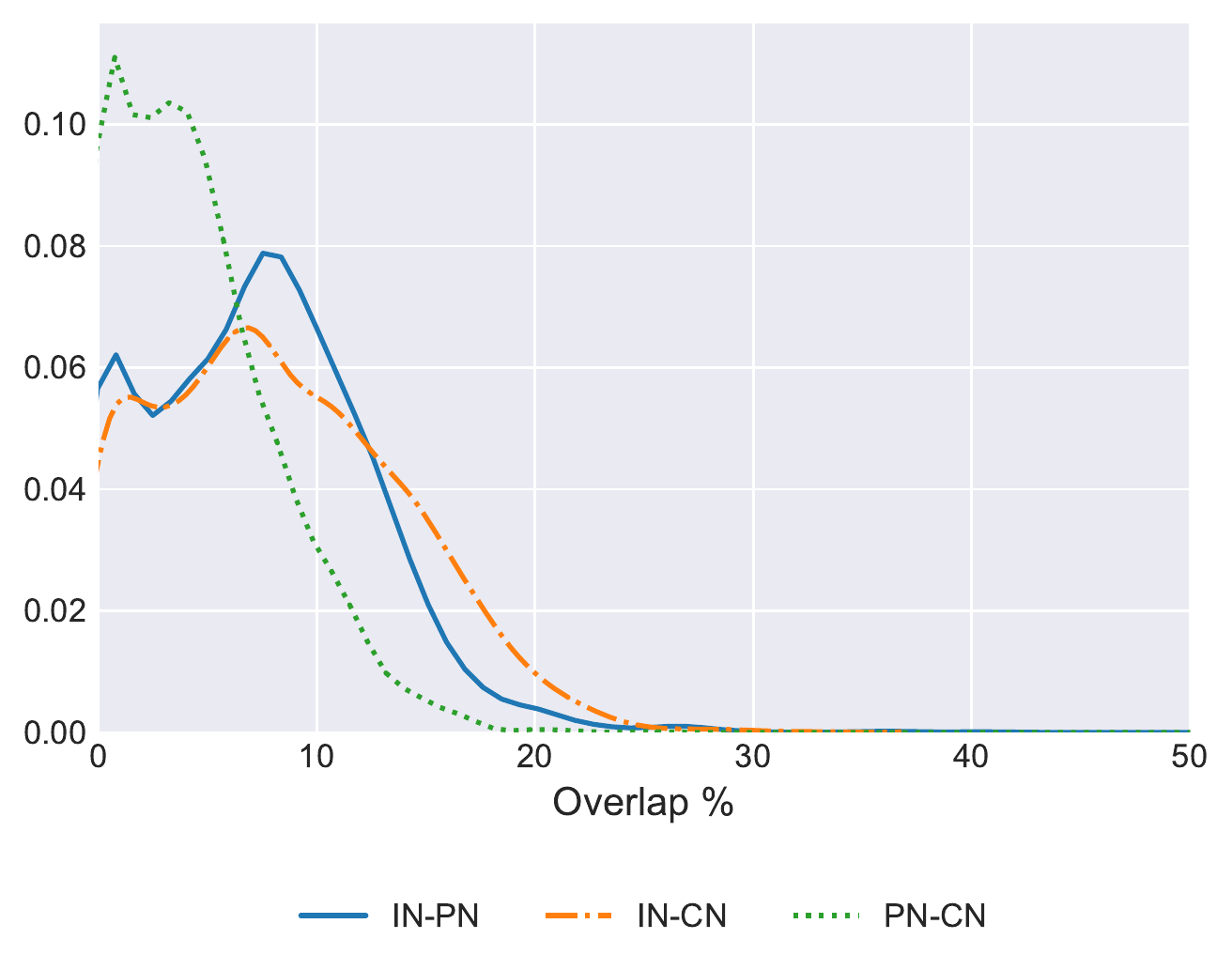}}
\subfloat[IN$_{DM}$ and PN$_{DM}$.
          \label{fig:subfig-b}]{\includegraphics{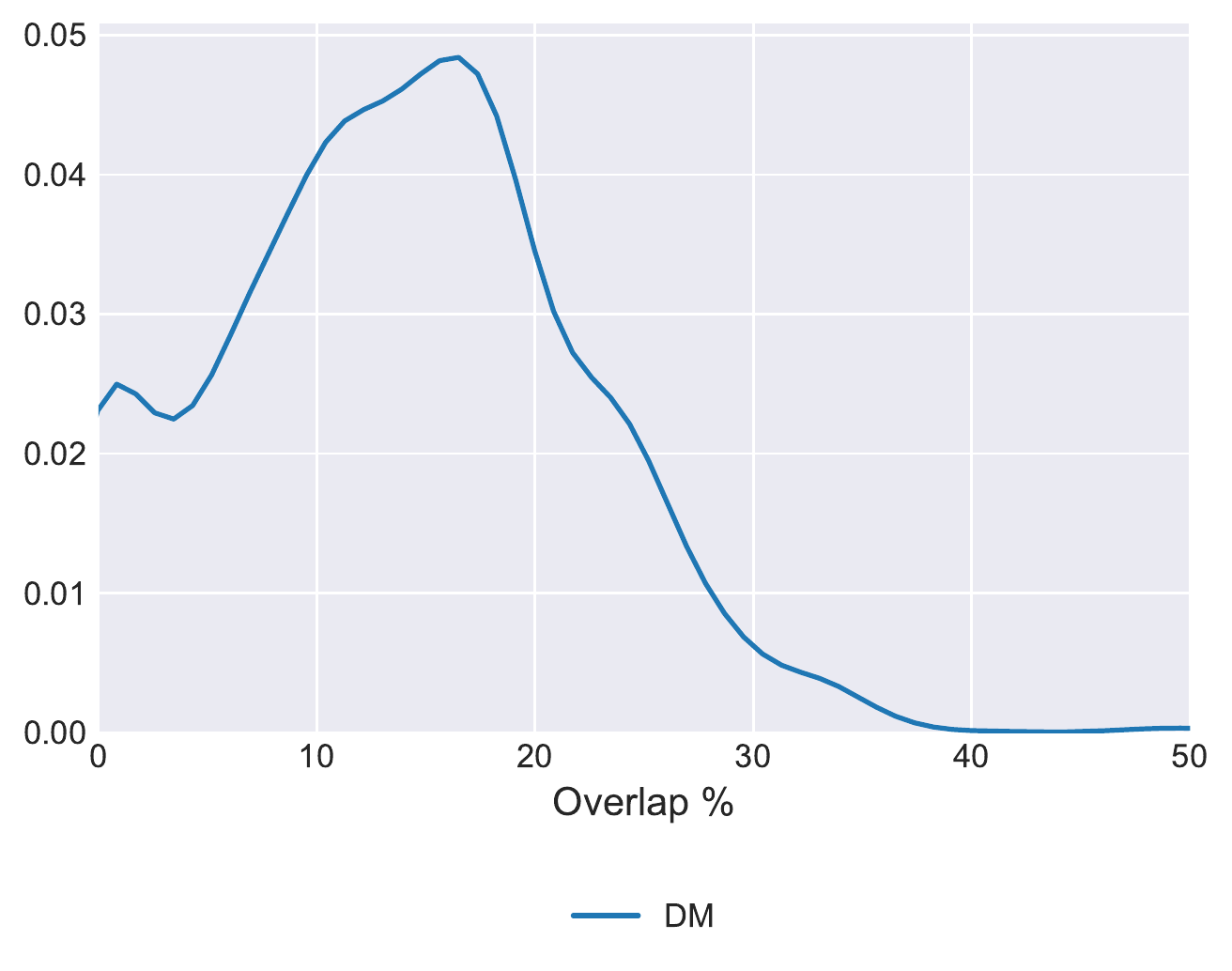}}
\caption{Similarity between CN, IN and DM in users dataset. }
\label{fig:common_similarity}
    \end{figure*}

For each user, we compute the similarity between their $IN_@$, $PN_@$, and $CN_{FR}$ features using Jaccard similarity, then we plot the distribution of the similarity score across all users. We repeat this process for the domains features by computing the similarity between $IN_{DM}$, $PN_{DM}$.
Figure \ref{fig:common_similarity} shows the similarity distribution between the network's sets, where zero indicates no overlap and 100\% means identical sets. We observe that there is a noticeable difference in each network for the same feature component. The overall similarity between accounts in each of the three networks ranges between zero and 20\%, and it ranges between 0 and 35\% for domains.
This result means that users tend to interact and like contents from users out side their connection network, and like tweets with links generally different from the domains they link in their tweets. 
This is actually an interesting finding, which actually raises further research question about the reason of having the performance of the three networks in stance detection similar when they are mostly different.

\begin{figure*}
    \centering
    \setkeys{Gin}{width=0.33\textwidth}
\subfloat[IN$_{@}$, PN$_{@}$ and CN$_{FR}$ - Favor.
          \label{fig:subfig-aa}]{\includegraphics{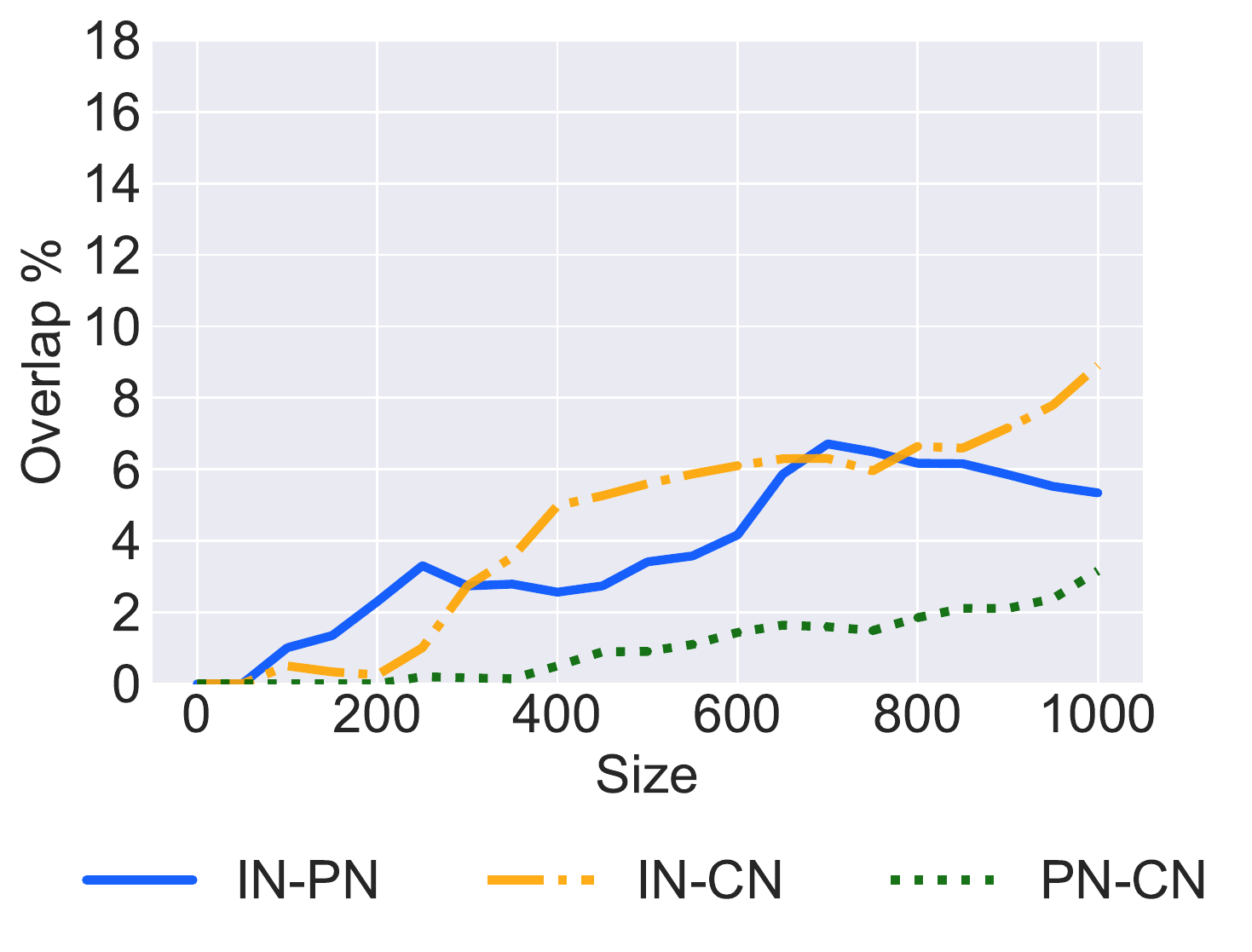}}
    \hfill
\subfloat[IN$_{@}$, PN$_{@}$ and CN$_{FR}$ - Against.
          \label{fig:subfig-bb}]{\includegraphics{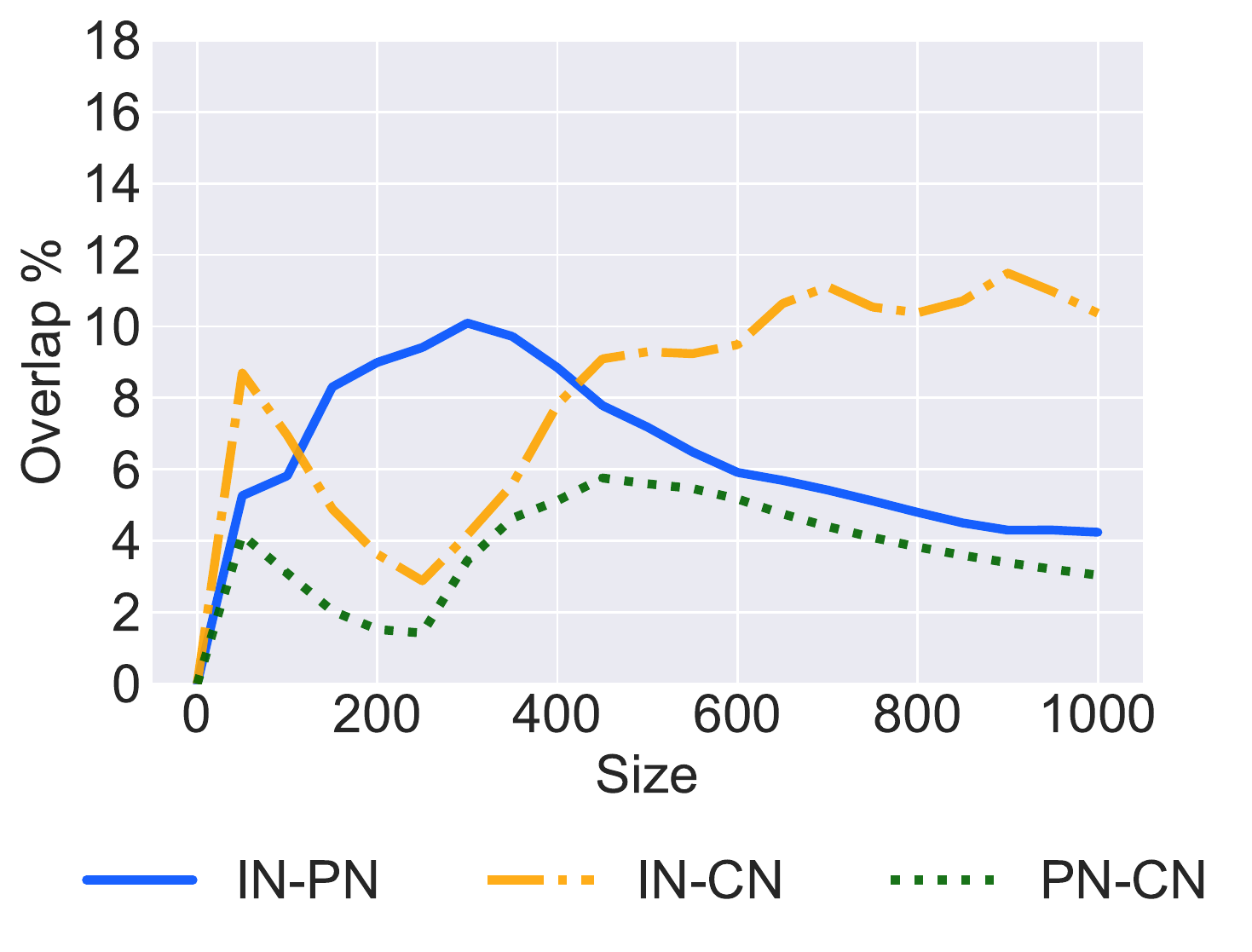}}
    \hfill
\subfloat[ IN$_{DM}$ and PN$_{DM}$.
          \label{fig:subfig-cc}]{\includegraphics{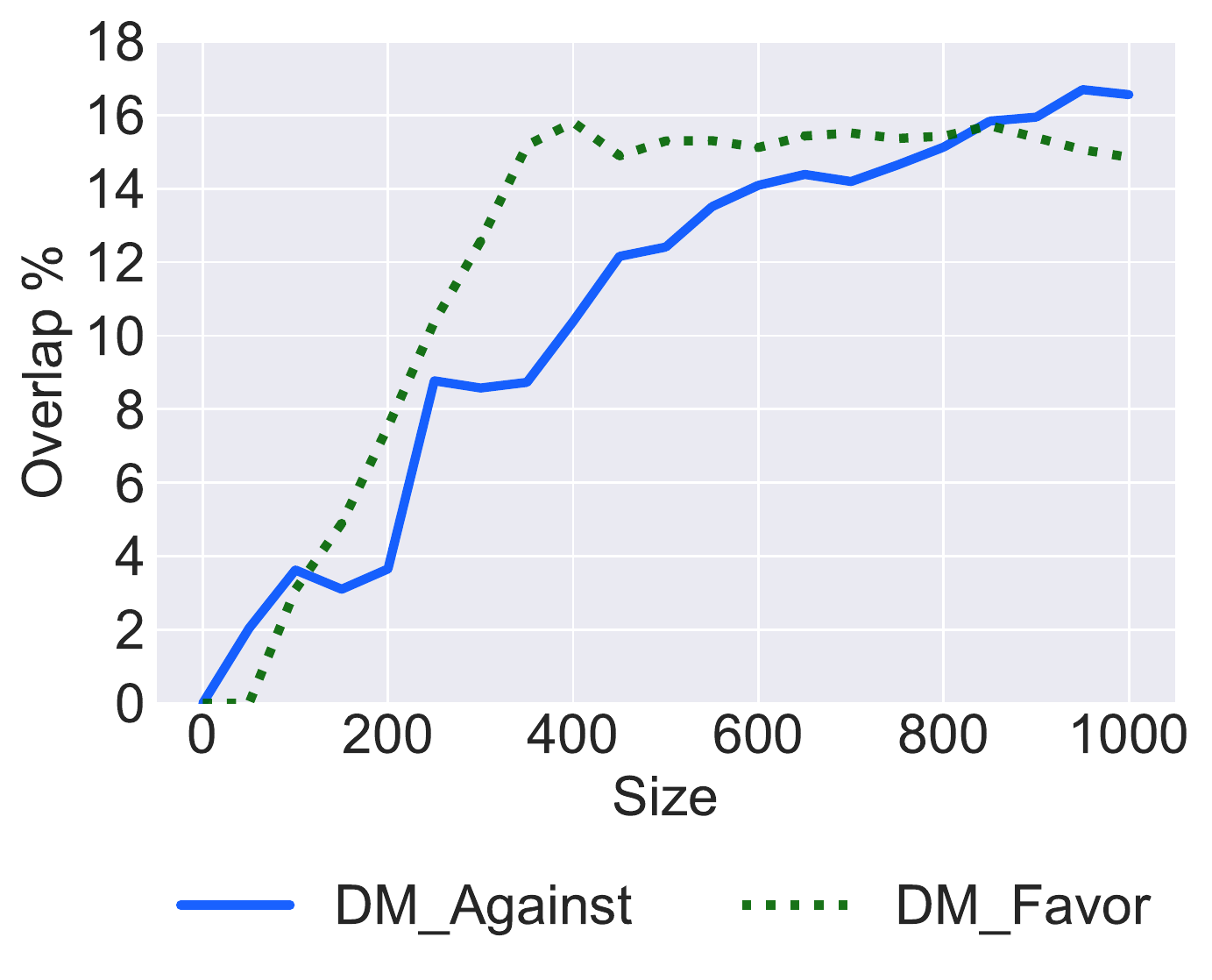}}
\hfill
\caption{Similarity between CN, IN and DM for (In-favor  and Against) stances with respect to the top features. }
\label{fig:common}
    \end{figure*}

There is a hypothesis behind the similar performance, that actually the small percentage of similar accounts (domains) between the three networks are those which create the most influential features for the classification, and thus the three classifiers achieved comparable performance. Therefore,
we further analyse the similarity between the most influential features of the three networks sets, where influential features are identified as those having the highest weights for each of the classes for each topic. We use Jaccard similarity to compute the similarity between the top $N$ influential features of IN, PN and CN and plot the similarity for $N$=\{1 $\rightarrow$ 1000\}. Figure \ref{fig:common} presents the similarity for each network features influencing favor and against stance.
Again, it is observed that similarity between the most influential features is not high for any of the networks for both `favor' and `against' classes, where the similarity does not exceed 10\% for the accounts, and 17\% for domains.

These findings confirm the differences between the three networks, and show that each network represents a different set of accounts and domains for the same user. Even with the most influential features for models trained on each classifier the set of features is different than the other. This means that a more in-depth analysis to these features is required to understand the high performance of the classifiers trained on the three fairly independent networks.

\subsection{Which Network Features Reveal the Stance?}

To get meaningful insights about the contribution of the features to infer the stance, we identify the most influential feature of the best model from (CN), (IN), and (PN) network with regards each topic. We hope this would give some explanation to the good performance of these models, especially after we found that these networks do not highly overlap. 

\begin{table}[t]
\centering
\footnotesize 
\begin{tabular}{ccll} 
  \hline 
  \textbf{T} & NW  & Favor & Against \\
  \hline
  
    & IN & @atheism\_tweets, @atheistrepublic, @god\_stupid & @ChristianInst, @godlesstheory,@godbiblechurch \\
A	& PN & @thetweetofgod,@foxnews,@nytimes  & @prayerbullets, @reuters, @cnn  \\
	& CN &  @Stephenfry, @RichardDawkins, @MarilynManson & @baptism\_saves, @srisri, @artofliving \\
\hline
	& IN & @telegraph, @independent, @climatereality & @skynewsbreak, @nytopinion, @reuters \\
CC	& PN & @nytstyles, @news4anthros, @fox2now & @cnn, @foxnews, @nythealth \\
    & CN & @barackobama, @potus, @mashable & @foxnews, @sentdcruz, @cnn \\
\hline

    & IN & @washtimes,@hillaryclinton,@realdonaldtrump &  @trumpstudents, @foxnewssunday,@brianschatz\\
HC	& PN & @cbsnews,@nbcnews, @hillaryfornh & @govchristie,@drbiden, @sentedcruz  \\
	& CN &@hillaryclinton, @billclinton,@shehasmyvote  &@foxnews, @realdonaldtrump, @madam\_presiden  \\
\hline
    & IN & @mtv,@goodreads,@feministculture & @feministfailure, @goodmenproject, @womenwriters \\
FM	& PN &@feministajones,@ppfa,@foxnews  & @nytopinion, @mtvnews,@weneedfeminlsm  \\
	& CN & @vday,@Schofe,@twitterfashion & @Truth\_seeeeker,@femalefedupwith,@thepowrhouse \\
\hline
    & IN & @humanesociety,skynews ,@ppactionca & @ppfa, @nbcnews, @bible\_time \\
LA	& PN &@savewomenslives,@dallasnews,@citynews  & @onejesusloves, @younglife, @yahoonews  \\
	& CN & @thedemocrats,@barackobama,@hillaryclinton  & @prolifeyouth,@march\_for\_life, @lifeteen\\
\hline
\end{tabular}
\caption{Top features extracted from the best model in each case and trained on two classes, CN$_{FR}$, IN$_{@}$,PN$_{@}$. }
\label{tab:results_topfeatures_2classes}
\end{table}
 
\begin{table}[t]
\centering
\footnotesize 
\begin{tabular}{ccll} 
  \hline 
  \textbf{T} & DM  & Favor & Against \\
  \hline
  
   & IN &sciencealert, thinkprogress, washingtonpost & nationalpost, washingtonpost, newsweek \\
A	& PN & reuters, newhumanist, telegraph & faithreel, bible,prayerbullets  \\
	
\hline
	& IN &thetimes, nytimesarts, nbcnews  & bbc, naturalnews, washingtontimes \\
CC	& PN &abc, newswire, nypost & cbc, telegraph, washingtontimes \\
    
\hline

    & IN &nytimes, thedailybeast, cnbc &opposingviews, washingtontimes, foxnews\\
HC	& PN &nytimes, theguardian, nbc & cnn, foxnews, newsfoxes \\
	
\hline
    & IN &cnn, buzzfeed, nytimes &dailymail, bbc, theguardian\\
FM	& PN &apnews, washingtontimes, feministing & independent, dailymail,activistpost  \\
	
\hline
    & IN & newstatesman, nytimes, cnn & nypost, dailymail, cbsnews\\
LA	& PN & bpas, ahealthblog, thenation & lifenews, gotquestions, cnsnews\\
	
\hline
\end{tabular}
\caption{Top features extracted from the best model in each case and trained on two classes,  IN$_{DM}$,PN$_{DM}$. }
\label{tab:results_topfeatures_2classes_DM}
\end{table}

Table \ref{tab:results_topfeatures_2classes} and \ref{tab:results_topfeatures_2classes_DM} show the top features that have a noticeable influence on the stance classification for each topic with respect to the weights of features in the linear SVM model for the best features from each network group:  (IN$_{@}$+IN$_{DM}$), (PN$_{@}$+PN$_{DM}$) and (CN$_{FR}$). 
 
In the (CN$_{FR}$) network, the social influence manifest through the users' friends (following network). Users tend to follow the accounts that support their stance. For instance, users with against stance toward legalisation of abortion (LA) tend to follow accounts that oppose the abortions such: '@prolifeyouth', '@march\_for\_life'. The same for the users with favor stance to Hillary Clinton where the top followers are '@Hillaryclinton', '@billclinton', '@shemyvote'. Users who have a favor stance towards Atheism tend to follow social actors with the same believes such:'@Stephenfry', '@RichardDawkins', '@MarilynManson'. Similarly, users with favor stance toward feminist movement follow the accounts that support feminism. One of the top features that identifies the in-favor stance toward feminism is '@vday', which is an activist movement account that supports the feminist movement as this account description indicates: ''to End Violence Against Women \& Girls.''. For the climate change and legislation of abortion, the politicians and news outlets are the most influential accounts in predicting the stance. We can not specify whether these users follow such account because they support  their opinion towards each topic.
 
Unlike CN, influential accounts for IN and PN include news accounts. For instance, the news accounts '@washtimes' and '@cbsnews' are one of the distinguishing features to detect the favor stance to Hillary Clinton in IN$_{@}$ and $PN_{@}$. 
In addition, '@telegraph' in $IN_{@}$  has a positive correlation with favor stance to climate change. Users with favor stance to the legalization of abortion interact with '@skynews' account. In contrast, news accounts have a minimal effect in detecting stance toward feminist movement and atheism, where the top mentions features that capture a favor stance are accounts that support the topic: '@atheism\_tweets' and '@feministcultur'. 

Also, another difference between IN and PN, is that IN usually contains accounts of opposing view since in this case the interaction can be through replying or quoted retweets with  opposing comments. This case can be seen in Hilary Clinton topic, where `@realDonalTrump' is one of the top features for the `favor' stance in IN. It can be imagined that the interaction here is not for support as shown in table \ref{tab:sample_tweets} (Example 3). In addition, interacting with accounts that have a related meaning to the topic seems to have a visible correlation with detecting the against stance of users. For instance, the interaction with   '@godlesstheory' and '@godbiblechurch' has an influence in detecting the against viewpoint to atheism.  Similarly, '@bible\_time' captures the against stance toward abortion. Furthermore, famous accounts with clear support to a related social issue have a clear influence in detecting the stance. For instance, users with against stance to feminism interact with '@feministfailure'. In addition, users who oppose the legalisation of abortion interact with '@ppfa', Planned Parenthood account. 


 
For the web domain features, it can be noticed that the top domains features $IN_{DM}$ and $PN_{DM}$ are mostly news websites. News websites and media outlets such as 'washingtonpost' and 'sciencealert' are one of the distinguishing features to detect favor stance toward Atheism.  In contrast to mentions, the news websites have a noticeable effect in detecting users view points toward feminist movements. We can see that users with against stance to feminist movement tend to share contents from 'dailymail', 'bbc' and 'theguardian' websites. Users with support stance to feminist movement tend to share contents from 'cnn'. Users with against stance to Hillary Clinton share contents from news websites such as 'opposingviews', 'washintontimes' and 'foxnews'. The website 'nytimes' has a positive effect in identifying the favor stance to Hillary Clinton. We can notice some overlap between $IN_{DM}$ and $PN_{DM}$, where it seems users like and interact with the same news and media outlets in the PN and IN networks. For instance, users with against stance to Hilary Clinton tweet interact and like news contents from 'foxnews'. The same for users with against stance to the feminist movement, the users like and interact with 'daily-mail'. In general, there is a tendency for the users to like and share content from the same media as described in the next section.

\begin{table}[t]
\footnotesize
\centering
\begin{tabular}{cccp{10.5cm}}
  \hline
 \textbf{\#}& \textbf{T}  &  \textbf{Feat}  & \textbf{Example tweets (favor)} \\
  \hline
1& CC &  IN$_{@}$ & RT @Telegraph: Prince Charles reveals his gardening inspiration: a hidden Buckingham Palace veg plot https://t.co/tBZB5DSKt5\\

2& HC &  PN$_{@}$ & @NBCNews Kill the bear for BEING A BEAR! What's wrong with this? \\

3& HC& IN$_{@}$& You are an idiot on so many levels,  @realDonaldTrump https://t.co/keptgYgTed\\

 4& FM  & IN$_{@}$ & I'm your nightmare come true," said Angela. \#YAlit \#vampire \#paranormal \#Action \#humor https://t.co/MCvYEvdz8Z @goodreads 	 
\\
  5& A & IN$_{@}$ & @god\_stupid @userid just the ignorant, racist, sexist, child abusing fanboys that roll play \#christianity.\#Atheist and proud  
\\

\hline
 \textbf{\#}& \textbf{T}  &  \textbf{Feat}  & \textbf{Example tweets (against)} \\
  \hline
 6& CC  &IN$_{@}$ &RT @SkyNewsBreak: Former Labour Prime Minister Tony Blair has told Sky News Theresa May will win the General Election \#GE2017 \\

7& A  & IN$_{@}$ & RT @ChristianInst: Romans 8:28 And we know that for those who love God all things work together for good, for those who are called accordi\\
  
8& A  & PN$_{@}$ & @prayerbullets: Turn every curse sent my way into a blessing -Neh. 13:2 \#Prayer\\
 
  \hline
\end{tabular}
\caption{Sample of tweets and the context of IN and PN in relation with stance and topic.}
\label{tab:sample_tweets}
\end{table}



\subsection{The Context of the Features}
 \label{se:similarity}


We carried a further qualitative analysis to identify the context in which the IN and PN features correlate with the topic of the target. Table \ref{tab:sample_tweets} shows a sample of tweets from the users' timelines (IN) and Favorite timeline (PN) with respect to topic-stance pair and highlights the interactive nature of the user with the top features. As explained in the previous section, what sets apart users with support/against stance to climate change are those pertaining to news portals. For instance, the most dominant mentioned accounts that influence the supporting and opposing position toward climate change is '@telegraph' and '@SkyNewsBreak'. Users interaction with these news accounts in the sense of re-tweeting and liking the news that has no relation to climate change (Example 1 and 6). 

Tweets from '@NBCNews' with no relevance to Hilary Clinton or the presidential candidates tend to be liked by users with a stance supporting Hillary, (Example 2). The same with users who support feminist movement, they interact with account '@goodreads' with no topical relation to stance topic, (Example 4).  The user mentioned @goodreads to promote to the novel "Beginnings" which is a
teen romance, sci-fi and fantasy story. Furthermore, example 6 demonstrates how the interaction
with @SkyNews helps in predicting the against stance towards Climate Change (CC) even with
news that does not concern with climate change.
In contrast, users opposing atheism tends to mention religious accounts to support their stance against atheism. For instance, users with against stance toward atheism interacted with '@ChristianInst' by retweeting verses from scripture (Example 7). Furthermore, users who have an against stance toward atheism tend to like religious's content from accounts such as '@prayerbullets' (Example 8). Users supporting atheism interact with accounts that are sarcastic toward religions such as  '@god\_stupid' account, in a sense of hashtag as a way of expressing the against viewpoint towards the religious people. The account '@god\_stupid' is a  sarcastic account,  yet the interaction with it tends to take a kind of attacking the religious means as shown in (Example  5). Similarly, Users supporting Hilary Clinton defending their viewpoint by attacking '@realdonaldtrump' (Example 3). 



\section{Discussion}

In this work, we studied the possible signals that can reveal the user's stance from their publicly available online data. Unlike most of the literature in this area, which mostly focuses on achieving a high accuracy without in-depth analysis, our main focus is to understand how stance could be revealed throughout different sets of signals. This led us to explore multiple sets of features including some that have not been examined before (such as the preference network), and test it on a stance benchmark dataset of multiple topics of different genres.

\subsection{What factors reveal stance and how?}
Our study investigates three main research questions that have not been sufficiently explored in earlier studies on stance detection.

Our first research question is concerned with exploring the different signals from user's public social media profiles that can reveal their stance. We have defined three sets of network features, including interaction (IN), preference (PN), and connection (CN) networks, and compared their performance to textual features that represent the state-of-the-art models on the SemEval dataset. Our findings showed that user's stance can be detected with many signals, including textual content and different sets of network features. 
We found that using network features leads to a more accurate stance detection than using content-based features solely, 
and the performance becomes statistically significantly better when both sets of features are combined together. We also noticed that when building a stance classifier, a binary classifier is more superior than a classifier that allows neutral stance, which could be linked to the argument that there is no ``neutral'' stance and everyone should have some leanings~\cite{jaffe2009stance}.

Our second research question focused on how the performance of the stance detection using these features would differ across different topics. Our analysis of the five topics in our dataset showed that network features consistently achieve better performance on average compared to textual features. We only found that the performance for one topic (CC) has always the lowest F score. Our investigation to the distribution of the stances on this topic suggests that the problem stems from the large imbalance in the training samples, which leads the prediction model to predict only the majority stance class, which is independent of the set of features used.

As for our third research question, which concerns with investigating what makes the introduced features effective for stance detection; we initially analyzed the overlap between the accounts and web domains for each of the users in our dataset in the three networks: IN, PN and CN to ensure that their similar performance is not the reason for their high similarity in their nodes. It was surprising to find them mostly dissimilar with low overlap among them with <20\% similarity between them.
This was interesting to see that each of them captures one side of the user's activity, and each can reveal their stance.
We further investigated the top features in each network model. We noticed that the top features can sometimes be topically unrelated to the target and yet have a high impact on deciding the stance of the topic.
For instance, the interactions with accounts as @goodreads and @SkyNews help in detecting the stance towards feminist movement (FM) and climate change (CC) respectively, as shown in section \ref{se:similarity}. Since these features have no direct relation to the topic of the stance, this indicates that the user's stance can be detected with many signals regardless of the topic.
We showed that using content-less features help in detecting the stance for the users with an implicit point of view toward a topic where the users may not directly express their point of view by using keywords related to the target. As the top features extracted from the two networks (PN$_{@}$) and (IN$_{@}$) have no direct relation to the stance's topic. For instance, the `@Telegraph' was one of the top features that predicts the in-Favor stance towards Climate Change topic.


Furthermore, one of the key findings from our study is the high performance of PN and CN for stance detection, which outperforms the state-of-the-art baseline TXT model. This shows that detecting stance for silent/passive users (who never tweet or share any content) is doable, given the condition that they have enough common signals in their preferences and connection networks.
This raises a real concern about the privacy of social media users in general, and motivates future research in the direction of protecting those users from having their leanings and beliefs revealed unconsciously~\cite{waniek2018hiding}.

Our experiments and analysis was applied to a set of five topics of political, social, and religious natures. Our findings show that regardless to the topic, there are usually signals in the users' online activity and connections that can reveal the stance of those users towards this topic. 

\subsection{Ethics and Privacy Considerations}
Using Twitter as a central platform for this study is supported by a robust  literature\cite{rajadesingan2014identifying,darwish_improved_2017,djemili2014does,mohammad_stance_2017}. Previous studies have shown that the stance detection in social media provides useful information to understand better the way in which people communicate and express their viewpoint towards a topic.  Stance detection has been used as the first step toward solving fake-news, polarization, and rumours ~\cite{ghanem2018stance,garimella2017reducing,zubiaga_discourse-aware_2018}. Hence, the famous Semeval stance detection benchmark dataset has been constructed using public tweets \cite{mohammad_semeval-2016_2016}. Furthermore, Twitter does not force demographic information of the user upon registration. Consequently, the  accounts are not linked to the physical identity of the users. In the process of collecting the additional tweets to extend the Semeval stance dataset we are using authorized developers accounts approved by Twitter application developer portal\cite{ahmed2017using}.   This study does not store non-public Twitter content, such as direct messages or other private or confidential information. The collected tweets are the publicly available data on Twitter as further indicated in the Twitter Developer Policy \footnote{\url{https://developer.twitter.com/en/developer-terms/agreement-and-policy.html}}.

\section{Conclusion and Future Work}
In this paper, we present a thorough analysis of the four main scenarios to predict the stance on social media. We investigate with a stance detection approach that can be text-independent, where the stances are predicted from users online activity. 
We introduce three sets of networks to represent users, which are the interaction, preference and connections network. The interaction network includes the accounts the user interacted with and the website domains the user shared; and the preference network represents the accounts and website domains in the tweets the user liked. Finally, the connection network is the set of friends and followers of the user. We conducted the experiments on SemEval 2016 stance benchmark dataset, and showed the superiority of network features over textual features when compared with the baseline model. All three network-based models outperformed the state-of-the-art methods that depend on textual features only. 
We also presented an analysis of the top features to identify the correlation between stance and topic with respect to the features groups. We explored with the key important features that have a positive effect on detecting stance for each target. The results denote accurate learning of the stances at the user-level representation that improves the content-related features model.

For future work, more analysis of the network features could be applied, since retweeted accounts might denote different preferences than replied or mentioned ones. In addition, it is essential to create new sets of data covering more topics to validate the generalisability of our findings. 
Finally, since our work raises some concerns about the vulnerability of users by having their stances easily detectable even when not directly discussing the topic, it becomes highly essential to develop methods to counter those automatic methods for detecting user's leanings as a step to protect user's privacy.

\bibliographystyle{ACM-Reference-Format}
\bibliography{ms}

\end{document}